\documentclass[twocolumn,prb]{revtex4}

\begin{document}


\title{Field-Induced Resistive Switching in Metal-Oxide Interfaces}

\author{S.~Tsui$^{1}$}
\author{A.~Baikalov$^{1}$}
\author{J.~Cmaidalka$^{1}$}
\author{Y.~Y.~Sun$^{1}$}
\author{Y.~Q.~Wang$^{1}$}
\author{Y.~Y.~Xue$^{1}$}
\author{C.~W.~Chu$^{1,2,3}$}
\author{L.~Chen$^{4}$}
\author{A.~J.~Jacobson$^{4}$}
\affiliation{$^{1}$Department of Physics and Texas Center for Superconductivity 
and Advanced Materials, 
University 
of Houston, 202 Houston Science Center, Houston, Texas 77204-5002}
\affiliation{$^{2}$Lawrence Berkeley National Laboratory, 1 Cyclotron Road, 
Berkeley, California 94720}
\affiliation{$^{3}$Hong Kong University of Science and Technology, 
Hong Kong}
\affiliation{$^{4}$Department of Chemistry, University of Houston, 
136 Fleming Building, Houston, Texas 77204-5003}

\date{\today}

\begin{abstract}
We investigate the polarity-dependent field-induced resistive switching phenomenon 
driven by electric pulses in perovskite oxides. Our data show that the switching 
is a common occurrence restricted to an interfacial layer between a deposited 
metal electrode and the oxide. We determine through impedance spectroscopy that 
the interfacial layer is no thicker than 10~nm and that the switch is accompanied 
by a small capacitance increase associated with charge accumulation. 
Based on interfacial $I-V$ characterization and measurement of the temperature 
dependence of the resistance, we propose that a field-created crystalline defect 
mechanism, which is controllable for devices, drives the switch.
\end{abstract}

\maketitle

Recent observation of room-temperature resistive switching driven by electric fields in 
various perovskite oxides has garnered attention due to the potential for nonvolatile memory 
applications. This field-induced resistive switch has been reported in several 
compounds.\cite{liu,bai,bec,wat1,tul1,tul2} The switching observed shares several features: 
a moderate switch speed; an altered resistance inconsistent with the well-accepted bulk 
resistivity; and a sensitivity to surface treatments. Various models have been put forth, 
but inconsistencies remain. For instance, bulk charge ordering in 
Pr$_{0.7}$Ca$_{0.3}$MnO$_{3}$ (PCMO) has been suggested, but this creates an apparent 
conflict with both the spatial symmetry and the reported bulk properties.\cite{liu} 
Although interface models have been proposed involving either lattice defects or carrier 
concentration,\cite{bai,tul1,ross} the exact nature of this interface remains vague. 
We therefore investigate the interface properties associated with the resistive switch and 
have observed: a) the switching is a common phenomenon to the metal-oxide interfacial layer; 
b) the interfacial transport properties are rather different from that of the bulk; 
c) the interfacial resistance is dominated by the carrier trapping with no indication of 
Schottky barriers; and d) the capacitance of the interfacial layer, on the order of 
1000 nF/cm$^{2}$, changes with switching, which is indicative of a change in the space-charge. 
These observations can be self-consistently understood using a carrier-trapping model and 
suggest that the field-induced switch is not restricted to a small class of materials. 
Therefore, the potential for controlling this phenomenon for device applications is 
considerable.

Ceramic samples synthesized by standard solid-state reactions were determined to be 
single-phase based on X-ray powder diffraction patterns taken on a Rigaku DMAX-IIIB 
diffractometer. PCMO thin films were ac sputtered on LaAlO$_{3}$ substrates at 
760~$^{\circ}$C in an Ar:O$_{2} = 2:3$ mixed atmosphere at 140~mTorr. The films were found to 
be highly epitaxial. Pt leads were attached to the ceramic samples using Ted Pella 
Leitsilber 200 Ag paint. The thin film samples received sputtered Ag electrodes to which Pt 
leads were also attached using Ag paint.

Ceramic LaCoO$_{3}$, La$_{0.7}$Ca$_{0.3}$MnO$_{3}$, Pr$_{0.7}$Ca$_{0.3}$MnO$_{3}$, 
SrFeO$_{2.7}$, RuSr$_{2}$GdCu$_{2}$O$_{3}$, and YBa$_{2}$Cu$_{3}$O$_{7}$ were tested. 
We adopted the multi-leads configuration developed by Baikalov \textit{et al.}\cite{bai} to 
measure both the interfacial and bulk resistances of the samples (inset, Fig.~\ref{fig:fig1}). 
A DEI PVX-4150 pulse generator provided 150~V pulses of 200~ns width. A pair of electrodes 
($A$ and $B$ in Fig.~\ref{fig:fig1}) at room temperature was exposed to twenty pulses 
spaced 3~s 
apart. As a typical example, data for LaCoO$_{3}$ (Fig.~\ref{fig:fig1}) revealed that 
the 4-leads 
bulk resistance $R_{bulk} \sim 100$~$\Omega$, which is consistent with previously published 
resistivity data,\cite{yam} remained unaffected by the pulses within an experimental 
resolution of 0.01~$\Omega$. However, the 3-leads interfacial resistances, $R_{A}$ and $R_{B}$ 
measured through the reference electrode $C$, showed hysteretic and polarity-dependent 
switching between a low (on)- and a high (off)-resistance state. The interfacial resistances 
at both states are much higher than that expected from the bulk value. All other tested 
compounds behaved similarly when exposed to pulsing. The data therefore suggest that the 
switching is a common phenomenon, but associated only with an insulating interfacial layer. 
In view of the different spin configurations, valences, and band structures of the tested 
compounds, the field-induced polarity-dependent resistive switch is unlikely to be associated 
with a particular phase transition or carrier doping.

In order to better characterize this interface, impedance spectroscopy was performed on PCMO 
thin films using a Schlumberg\"er SI 1260 impedance analyzer with a Solartron 12603 
attachment. Measurements spanned a frequency range of 0.1~Hz to 20~MHz. The single-layered 
PCMO strips received three parallel $0.2 \times 1.3$~mm$^{2}$ Ag electrodes spaced 0.3~mm 
apart. The electrodes $A$ and $B$ experienced switching under the same 150~V, 200~ns wide 
pulses. Electrode $C$ served as a reference electrode. Both the 2-leads impedances $Z_{AB}$ 
and $Z_{BC}$ were then measured. In general, the impedance can be modeled as three parallel 
RC circuits associated with the two interface regions and the bulk portion 
(inset, Fig.~\ref{fig:fig2}). Therefore, we expect the impedance spectrum to show three 
sections to account for our model, assuming that the corresponding time constants differ 
(Fig.~\ref{fig:fig2}).\cite{sin}

Indeed, $Z_{AB}$ and $Z_{BC}$ share a three-sectioned feature. The impedance spectra show two 
merged semi-circles that begin on the $Z'$ axis. Below 1~kHz, the spectra terminate on the 
$Z'$ axis at values that correspond to the total 2-leads dc resistance measured between the 
electrodes. Above 1~MHz, the spectrum again reaches the $Z'$ axis at a left endpoint with a 
resistance in rough agreement with the bulk resistance measured in the 4-leads configuration. 
Therefore, the $Z'$ segment between the origin and the endpoint is attributed to the bulk 
portion of the sample based on the negligible $C_{AB}$ and $C_{BC}$ estimated. It should be 
noted that the positions of the left endpoints are unaffected by the switching, which is 
a further indication of the bulk contribution. At the intermediate frequencies, both spectra 
are dominated by a large semicircle. In the case of $Z_{AB}$, it represents a parallel RC 
circuit of 700~$\Omega/1.6$~nF and 1500~$\Omega/2.2$~nF for the on- and off-states, 
respectively. We assign this feature to electrode $B$ based on the measured dc interfacial 
resistances. Similar results were obtained for $Z_{BC}$ with some difference in the $R_{B}$ 
and $C_{B}$ deduced. We attribute the differences to inhomogeneity along the width of 
electrode $B$. An additional small segment appearing as a merged semicircle between these two 
major features is attributed to electrodes $A$ and $C$, respectively. The interpretation is 
supported by the fact that the $R_{C}$ and $C_{C}$ deduced are not altered by switching, and 
the three-leads $R_{A}$ changes only slightly with switching.

We expanded upon the spectroscopy data with a capacitance measurement across a multilayered 
Ag/PCMO/Pt sandwich similar to the configuration used by Liu \textit{et al.}\cite{liu} In 
order to properly measure the capacitance of the interface, bypass current through the Pt 
bottom layer had to be prevented. This was achieved by making the Pt layer ultrathin with 
poor in-layer conductance, creating an overall resistance $\sim 10$~k$\Omega$ between a 
switched 
electrode and a reference electrode. As a result, we were able to directly measure the 
interfacial 1.5~nF capacitance, a value consistent with the above spectrum analysis, using 
the 3-leads configuration with a HP 4285A LCR meter.

Knowing the capacitance $C$, the thickness of the interfacial layer, 
$d = \epsilon_{0}\epsilon S/C$, can be estimated, where $\epsilon_{0}$, $\epsilon$, and $S$ 
are the vacuum permittivity, the dielectric constant, and the electrode area, respectively. 
With $\epsilon_{0} = 8.85 \times 10^{-14}$~F/cm, $C \sim 2 \times 10^{-9}$~F, and 
$S = 0.02 \times 0.13$~cm$^{2}$, a $d$ is obtained to be 1 and 10~nm with $\epsilon = 1$ and 
10, respectively.\cite{note} Thus, the switching layer seems to involve no more than 30~unit 
cells in the thickness.

To further characterize the switching mechanism, we investigated the temperature dependence 
of the resistance and the $I-V$ characteristics of the interface. The temperature 
dependences of the interfacial resistances of the PCMO ceramic at both on- and off-states 
were found to be identical (dashed lines, inset, Fig.~\ref{fig:fig3}), demonstrating that the 
hopping barrier, hence the mobility, is the same in both states. However, the temperature 
dependence of the bulk (solid line, inset, Fig.~\ref{fig:fig3}) differs visibly from that of 
the interfaces, suggesting that the interfacial layers are significantly altered.

Carrier transport in insulators and semiconductors has been extensively studied. The 
dominant factor will either be space-charge limited currents (SCLC), in which the carrier 
injection from the electrode is limited by the electric field of the accumulated space 
charges, or the carrier creation through thermionic emission limited conduction (TELC), 
electrothermal Poole-Frenkel emission, or Fowler-Nordheim quantum tunneling. These modes can 
be distinguished via the isothermal $I-V$ correlation: linear $I \propto V$ for TELC; 
exponential $\ln I \propto V$ for Poole-Frenkel or Fowler-Nordheim emission; and 
$I \propto V^{2}$ for SCLC.\cite{wat2} Asymmetry in the $I-V$ polarities would suggest 
additional Schottky barriers.

The $I-V$ characteristics obtained for a Ag/ceramic PCMO interfacial layer 
(Fig.~\ref{fig:fig3}) show linear behavior up to approximately 0.1~V, suggesting that TELC is 
the dominant mechanism at the lower voltage regime where carrier injection plays a minor role. 
Beyond that, the behavior evolves into SCLC, as determined by the fit $I \propto V^{2.3}$ 
above 0.5~V. In this upper regime, carrier injection from the electrodes plays a dominant 
role. It is interesting to note that the resistance ratio in both regions is roughly 
voltage-independent. The data also show no appreciable asymmetry with the $I-V$ polarity 
within $\pm 1$~V, so the role of Schottky barriers appears to be minor, if it exists at all.

Analysis of the SCLC regime can be carried out using widely accepted models. Rose\cite{rose} 
and Lampert\cite{lam} obtained

\begin{eqnarray}
I = (\epsilon\mu_{0}N_{c}e^{-E/kT})V^{2}/(N_{t}t^{3})
\label{eq:one},
\end{eqnarray}
where $\epsilon$ is the dielectric constant, $\mu_{0}$ is the free carrier drift mobility, 
$N_{c}$ is the effective density of states in the valence band, $E$ is the effective trapping 
potential, $N_{t}$ is the number of shallow traps, $T$ is temperature, $V$ is applied voltage, 
and $t$ is the distance between electrodes. In particular, the doping level (or the 
carrier-creation rate) will not affect transport in SCLC, in which the dominant carriers are 
injected from the electrodes. The remaining factors will be the mobility and the fraction of 
free charge $\theta$ given by

\begin{eqnarray}
\theta = (N_{c}/N_{t})e^{-E/kT}
\label{eq:two}.
\end{eqnarray}
Both parameters will alter the resistance. However, significant changes in $E$ or $\mu_{0}$ 
contend with the identical temperature dependences of the on- and off-states 
(inset, Fig.~\ref{fig:fig3}). On the other hand, the pulse field of $10^{7}$~V/cm is close to 
the decomposition threshold and can conceivably create dense crystalline defects through 
electromigration. We consequently propose that a pulse-driven reversible change of the trap 
density, roughly a 5--100$\times$ variation, is responsible for the switching. The variation 
of $C_{B}$ with switching (Fig.~\ref{fig:fig2}) suggests a change in the space-charge 
distribution and supports the model. It should be noted that the same carrier trapping will 
increase the resistance in the TELC regime by the same factor, in agreement with our data.

In conclusion, we demonstrate that field-induced resistive switching is a common recurring 
phenomenon restricted to an interfacial layer between a metal and a perovskite oxide. We also 
characterize the interfacial layer responsible for the switching as an insulating region no 
thicker than 10~nm. Further characterization through $I-V$ measurements and the temperature 
dependence of resistance enables us to propose a conduction mechanism dominated by 
pulse-generated crystalline defects. The results obtained in the present study suggest that 
the switching is not compound-specific and can be optimized for device applications by 
modifying the interface.

\begin{acknowledgments}
The work in Houston is supported in part by 
NSF Grant No. DMR-9804325, the T.~L.~L. Temple Foundation, the John J. and Rebecca 
Moores Endowment, the Robert A. Welch Foundation, and the State of Texas 
through the Texas Center for 
Superconductivity at the University of Houston; and at Lawrence Berkeley 
Laboratory by the Director, Office of Science, Office of Basic Energy Sciences, 
Division of Materials Sciences and Engineering of the U.S. Department of Energy 
under Contract No. DE-AC03-76SF00098.
\end{acknowledgments}

\begin{figure}
\caption{\label{fig:fig1}Simultaneous switching of two electrodes on ceramic 
LaCoO$_{3}$ using twenty 150~V pulses of 200~ns width and 3~s apart. After pulsing, 
ten data points were measured. This was followed by another session of pulsing, 
after which another ten points were measured, as demarcated by the sequential count. 
The bold plot at the bottom shows the 4-leads bulk resistance. The open triangle plot 
shows the 3-leads interfacial resistance $R_{A}$, and the open circle plot shows 
$R_{B}$. This data is typical of the perovskite oxides we tested.}
\caption{\label{fig:fig2}Impedance spectra for a PCMO thin-film strip. 
Squares represent measurements over $Z_{BC}$ whereas circles represent $Z_{AB}$. 
Open and solid symbols represent the on- and off-states, respectively.}
\caption{\label{fig:fig3}$I-V$ data show TELC behavior in the lower voltage regime 
and SCLC behavior in the higher regime with $I \propto V^{2.3}$. The inset shows a 
visible difference in the temperature dependence between the interfacial and bulk 
resistances.}
\end{figure}


\begin{thebibliography}{99}
\bibitem{note} The upper limit of the bulk PCMO $\epsilon$ can be estimated as 
$\ll$ 10 through 
the observed capacitance across an Ag/PCMO/Pt multilayer film by using the film 
thickness as $d$. We tentatively take this value to be the upper limit of the 
interfacial $\epsilon$.
\end{thebibliography}

\end{document}